\documentstyle[aps,preprint]{revtex}
\tightenlines
\begin{document}
\draft
\title{Statistical and Dynamical Fluctuations in Heavy Ion Collisions: Role of Conservation Laws in Event-by-Event Analysis}
\author{G. Odyniec}
\address{{
Lawrence Berkeley National Laboratory, University of California, Berkeley, CA 94720, USA.\\
}}
\date{\today}
\maketitle
\begin{abstract}
%
The analysis of the statistical and dynamical fluctuations in nucleus-nucleus collisions on an event-by-event basis strongly relies on a comparison with specially constructed artificial events where statistical fluctuations and kinematical correlations are under control. In this paper, we present a novel, analytical method of constructing reference events based on independent emission, modified by the energy/momentum constraint, which can lead to a better understanding of the nature of the observed final-state fluctuations. This approach can be easily used in the analysis of other topics in the heavy ion field  (e.g. flow, HBT etc.) allowing more precise measurements.
\end{abstract}
%
%
\vspace{1.0cm}
An event-by-event analysis (e-b-e), so successfully used from the very beginning of high energy physics (bubble and streamer chamber experiments), was recently proposed \cite{Gaz} and applied \cite{Rol} to ultra-relativistic heavy nucleus-nucleus collisions at the CERN SPS energy. In hadronic physics, the e-b-e results provided information on the properties of individual interactions and their variations from event to event. In the heavy ion field, the search for ``unusual'' events, i.e. events having a particularly high variation of some variable from its average value, is especially important due to the expectation of non-trivial dynamical fluctuations caused by the formation of Quark-Gluon Plasma (QGP) bubbles and/or other exotic phenomena, such as a disoriented chiral condensate (DCC)\cite{Raj}, jet quenching \cite{Wan}, color fluctuations in the early stages of the collision \cite{Mro} and others. Recent data from the CERN SPS indicate that the energy density reached in nucleus-nucleus collisions has already exceeded the estimated critical value for QGP formation in the ``average'' event, thus the e-b-e approach may allow one to separate events in which plasma was created from those in which it was not.
Futhermore, it may allow one to determine to what degree the transient QGP phase and the dynamics of the phase transition to hadronic matter affect fluctuations observed in the hadronic state at freeze-out. 

The obvious, necessary experimental condition for e-b-e analysis is a large phase-space acceptance. However, even having a large fraction of the charged particles under control, one is confronted with the uncertainties related to the neutral, usually undetected, particles produced in the collision.
Therefore, it is essential to determine to what degree the fluctuations present in the experimental data (from the measurable part of the phase space) represent those of the entire event (relevant for comparison with the theoretical predictions).

In this paper we outline a novel, analytical method of deriving complete information on the entire reference (so called ``mixed'') event, based exclusively on the available experimental information - the fraction of final-state particles observed in the detectors - and the central limit theorem. Our approach, as described below, substantially refines the traditional one, where mixed events were composed by drawing particles randomly from the huge pool created by combining large amount of the data from the same trigger. Most importantly, we impose constraints from the conservation laws with associated kinematical correlations; this to the best of our knowledge, has never been done before.
Some aspects of this concept have been already applied in our earlier work (flow analysis of the Bevalac streamer chamber data) \cite{Dan}.

In order to establish whether the observed fluctuations are partly dynamical in  nature, we need to disentangle statistical effects i.e. effects due to the finite number of particles in the final state of the collision. In the following we will concentrate on the transverse momentum distribution to demonstrate the method qualitatively; it can also be applied to other observables.

First, we define a scale for measuring fluctuations and subtracting the trivial, statistical effects from the overall event-to-event variations. We start by comparing the width of the experimental spectrum with that of the specially constructed, mixed events of the same multiplicity for which we assumed independent particle emission, modified by the momentum/energy (p$_x$,p$_y$,p$_z$,E) conservation laws\footnote{The momentum/energy conservation explicitly addresses correlations particularly related to the observable p$_t$ chosen for our example. To investigate different observables, other constraints need to be included: e.g. the $K/\pi$ ratio requires the imposition of strangeness conservation.}. In mixed events, each particle is sampled from a different heavy ion collision belonging to the same data set (data taken with the same trigger)\footnote{If N is large - mixed and real (data) events have the same single-particle spectra; if N is small - one needs to apply a correction factor of order (N-1)/N.}.

\[
\rho^{\rm{mix}}_{_N} (p_{_1}, p_{_2}, ...) = \prod_{i=1}^{_N} \rho_{_1}^i(p_{i})\,\delta (P-\Sigma\, p_{i})
\]

\noindent 
where

$\rho^{\rm{mix}}_{_N}$ - density of all (N) particles of the mixed event

$p_{_1}$, $p_{_2}$, ... - particle momenta

$\rho_{_1}^i(p_{i})$ - density of a single particle having momentum p$_i$ in the i-th event

$\delta (P-\Sigma\, p_{i})$ - imposes momentum/energy conservation

P - total momentum/energy of the initial state = P$_{beam}$ in a fixed-target experiment

$\Sigma\, p_{i}$ - total momentum of all particles in the final state. 

\noindent
While dynamical fluctuations were totally eliminated from the mixed events during construction by the independent emission mechanism, we carefully preserved their exact multiplicity: i.e., for each data event of multiplicity N, we took one particle from N different data events with the same trigger to compose an equivalent mixed event of N particles.  

The width of the event distribution constructed thusly provides a benchmark for our analysis. It represents the scale of the statistical fluctuations and a measure of the probability for events to appear in the tails of the spectrum (rare events), where the fluctuations are largest.

The most important advantages of this strategy include:
\begin{itemize}
\item
model independence
\item
the same experimental systematics (efficiencies, acceptances, resolutions, etc.) in data and mixed events 
\item
consistency with the conservation laws (via the $\delta$ function), not only excluding ``non-physical'' events from the mixed-events sample, but also introducing into mixed events the kinematical correlations naturally present in the data.
\end{itemize}

\noindent
We use the following notation:

each event consists of N particles

1, 2, 3, ..., m  denote the charged particles observed in the experiment

m+1, m+2, ...., N denote the non-observed particles (neutral, outside the acceptance, etc.)

We express the density of observed particles as the density of all particles integrated over the unobserved part of phase space using the central limit theorem (for simplicity we drop the superscript ``mix'' from the left-hand side):

\[
\rho_{_m} (p_{_1}, p_{_2}, ...p_{_m}) = \int \rho_{_N}(p_{_1},p_{_2},...p_{_m},p_{_{m+1}},... p_{_N})\, dp_{_{m+1}}...dp_{_{N}}=
\]

\noindent
Next, we replace the density of all particles $\rho_{_N}$ by the superposition of independent single particle densities (assuming independent emission) and separate observed and unobserved particles. We also write down momentum/energy conservation explicitly.

\[
= \int \prod_{_{k=1}}^{_m} \rho_{_1}^{_k}(p_{_k}) \prod_{_{k=m+1}}^{_N} \rho_{_1}^{_k} (p_{_k})\:\delta ( P-\sum_{_{i=1}}^{_m}p_i - \sum_{_{m+1}}^{_N}p_i)\:dp_{_{m+1}}...dp_{_{N}}=
\]
\noindent
The product of the densities of all unobserved particles 
\mbox{$\displaystyle \prod_{_{k=m+1}}^{_N} \rho_{_1}^{_k} (p_{_k})$}
is substituted by one random variable 
\mbox{$\displaystyle\tilde{\rho} (\sum_{_{m+1}}^{_N}p_k)$}
representing the total missing momentum in the collision. Note, that
\mbox{$\displaystyle\int \prod_{_{k=m+1}}^{_N} \rho_{_1}^{_k}(p_{_k})dp...\equiv \int\tilde{\rho} (\Sigma_{_{m+1}}^{_N}p_k)d(\Sigma p_k)$}. 
The new variable 
\mbox{$\displaystyle\tilde{\rho} (\sum_{_{m+1}}^{_N}p_k)$} reflects much better the measuring capabilities of the experiment.

Multiplicity at CERN SPS energies is high, ranging up to a few thousand particles in central Pb+Pb collisions; therefore, k is large enough to allow us to use the central limit theorem to approximate the density of unobserved particles by a normal distribution $\cal N$: 
\[
{\tilde{\rho}(\sum_{_{k=m+1}}^{_N}p_k)\approx
\cal N} (\sum_{_{k=m+1}}^{_N}p_{_k}) 
\]
where
\[
{\cal N}(y) = \rho \cdot e^{{-\frac{1}{2}} A_{_{ik}}(y^i-\langle y \rangle^i) (y^k-\langle y \rangle^k)}.
\]
\noindent
$A_{ik}$ is the inverse of the covariance matrix 
\[
A = C^{-1}
\]
\[
 C\equiv cov(y_i,y_k) = (y^i-\langle y \rangle^i) (y^k-\langle y \rangle^k)
\]
and i, k = 1, 2, 3, 4 $\;\;$ (e.g., $p_x, p_y, p_z,$ E).

\noindent
We assumed that all errors are the same, therefore, all weights are the same.
\[
= \prod_{_k=1}^{_m}\rho^{_k}(p_{_k})\int {\cal N}(\sum_{_{k=m+1}}^{_N}p_{_k})\delta ( P-\sum_{_{k=1}}^{_m}p_k - \sum_{_{k=m+1}}^{_N}p_k)\:d(\sum_{_{k=m+1}}^{_N}p_k) =
\]
\noindent
after integrating with the $\delta$ function, we obtain
\[
= \prod_{_k=1}^{_m}\rho^{_k}(p_{_k})\ {\cal N}(P-\sum_{_{k=1}}^{_m}p_k).
\]
\noindent

Note, that by using the central limit theorem and by integrating with the $\delta$ function we were able to eliminate all variables related to experimentally inaccessible particles. The final results:

\[
\rho_{_m} (p_{_1}, p_{_2}, ...p_{_m}) = \prod_{_k=1}^{_m}\rho^{_k}(p_{_k})\ {\cal N}(P-\sum_{_{k=1}}^{_m}p_k)
\]
 depend only on the single-particle quantities measured in the experiment and on the weighting factor, also totally calculable from experimental data.
Thus, our task narrows down to computing the weighting factor
\mbox{$\displaystyle{\cal W}={\cal N}(P-\sum_{_{k=1}}^{_m}p_k)$}.
\noindent

Before we discuss the practical aspects of computing $\cal W$, let us make two digressions on statistics:

\begin{itemize}
\item
${\cal W}$ has a number of very convenient features:

\begin{itemize}
\item
${\cal W}\in$ (0,1)
\item
it eliminates the events where momentum/energy are poorly conserved
\item
it treats properly correlations resulting from conservation laws
\item
it is totally known from the experiment
\end{itemize}

\item
The covariance matrices for the sum over the observed particles and for the sum over the unobserved particles are equal.
\end{itemize}

Having sketched a general outline of the new method, let us point out the necessary steps to calculate weights ($\cal W$) and to construct proper reference events.

\noindent
The entire procedure is factorized to the five steps:

\noindent
step 1: For all observable particles in each event calculate 
\mbox{$\displaystyle W = \sum_{_{n=1}}^{_m}p_k$} where each $p_k$ is a momentum-energy four vector.

\noindent
step 2: Calculate the averages $\langle W^i \rangle$ and elements of the covariant matrix $c_{ik}$ for the entire set of events\footnote{Very schematic. Full procedure is more complicated.}:
\[
c_{_{ik}} = \langle ( W^i - \langle W^i \rangle)( W^k -\langle W^k \rangle )\rangle 
\]
where W is a four vector and i,k denote its components.

\noindent
step 3: Calculate the inverse of the covariance matrix: A = C $^{^-1}$ $\:$ (A$\cdot$C = I).

\noindent
step 4: Find $\cal W$ $\:(\equiv\cal N({\rm W}$) for each event
\[
{\cal N} = {\rm const} \cdot e^{{-\frac{1}{2}}(W-\langle W \rangle)^TA (W-\langle W\rangle)} \;\;\;\; (W\equiv {\rm four\:vector})
\]
\noindent
to be used in: 

\[
\rho_{_m} (p_{_1}, p_{_2}, ...p_{_m}) = \prod_{_k=1}^{_m}\rho^{_k}(p_{_k})\cdot \cal {N}({\rm W}). 
\]

\noindent
step 5: The last step we call ``simulations of measurements'':
Let us assume that our calculations show that some particular ``mixed'' event has the probability $\cal N$(W) of 0.3. We need to convert this number to 1 (= event entering our reference data sample) or 0 (= event rejected) to have a uniform treatment with the experiment (all events collected on the DST have probability = 1; all missing, of course, have probability = 0). This conversion is done in the following way: we draw a random number ($\alpha$) between 0 and 1, and compare our calculated weight $\cal W$=$\cal N$(W) against it.

If $\alpha$ is smaller than the $\cal W$ of the particular event, this event is accepted with a new probability = 1; however if $\alpha$ is bigger than $\cal W$ - the event does not enter our sample of reference data. So, in our example, the randomly selected $\alpha$ has to be smaller than 0.3 in order for the event to be accepted into the reference data sample.

The method described above is presently being tested with both Monte Carlo and experimental data. The quantitative understanding and evaluation of the results will take some time. In particular, in order to precisely reproduce the single particle distribution we have to introduce an additional weighting factor which compensates for the exponent resulting from the integration with the $\delta$ function. However, while still in the process of testing, we would like to communicate and make it available to the community, due to its wide range of applications for topics other than fluctuations analysis. We expect to present results from our simulations and data analysis soon. \\
Acknowledgements: I thank Pawel Danielewicz with whom the ideas presented here were developed. This work was supported by the Director, Office of Energy Rese
arch, 
Division of Nuclear Physics of the Office of High Energy and Nuclear Physics 
of the US Department of Energy under Contract DE-AC03-76SF00098.
\nopagebreak

%
%
%
%
\end{document}